          [2001/04/25 1.1 (PWD)]
\documentclass[a4paper,twocolumn]{esapub} 
\usepackage{natbib}
\usepackage{epsfig}

\title{MGGPOD: a Monte Carlo Suite for Gamma-Ray Astronomy}

\author[1,2\footnote{ESA Fellow}]{G. Weidenspointner}
\author[1,2]{M. Harris}
\author[3]{C. Ferguson}
\author[2]{S. Sturner}
\author[4]{B. Teegarden}
\author[5]{C.~Wunderer}
\affil[1]{Centre d'{\'E}tude Spatiale des Rayonnements, BP 4346, 31028
Toulouse Cedex 4, France} 
\affil[2]{NASA Goddard Space Flight Center, Code 661 \& USRA,
Greenbelt, MD 20771, USA}
\affil[3]{University of Southampton, Southampton, SO17 1BJ, UK}
\affil[4]{NASA Goddard Space Flight Center, Code 661, Greenbelt, MD
20771, USA}
\affil[5]{Space Science Laboratory, University of California Berkeley,
CA 94720, USA}

\begin{document}


\maketitle

\begin{abstract}
We have developed MGGPOD, a user-friendly suite of Monte Carlo codes
built around the widely used GEANT (Version 3.21) package. The MGGPOD
Monte Carlo suite and documentation are publicly available for
download. MGGPOD is an ideal tool for supporting the various stages of
gamma-ray astronomy missions, ranging from the design, development,
and performance prediction through calibration and response generation to
data reduction. In particular, MGGPOD is capable of simulating {\it ab
initio} the physical processes relevant for the production of
instrumental backgrounds. These include the build-up and delayed decay
of radioactive isotopes as well as the prompt de-excitation of excited
nuclei, both of which give rise to a plethora of instrumental
gamma-ray background lines in addition to continuum backgrounds.
\end{abstract}


\section{Introduction}
\label{intro}

Intense and complex instrumental backgrounds, against which the much
smaller signals from celestial sources have to be discerned, are a
notorious problem for low and intermediate energy gamma-ray astronomy
($\sim$50~keV -- 10~MeV). Therefore a detailed qualitative and
quantitative understanding of instrumental line and continuum
backgrounds is crucial for most stages of gamma-ray astronomy
missions, ranging from the design and development of new
instrumentation through performance prediction to data reduction.  A
promising approach for obtaining quantitative estimates of instrumental
backgrounds is {\sl ab initio} Monte Carlo simulation \citep[see
e.g.][]{Dean03}. 

We have developed a suite of Monte Carlo packages, named MGGPOD
\citep{weidenspointner_mggpod_seeon, weidenspointner_mggpod}, that
supports this type of simulation. The MGGPOD Monte Carlo suite
(version 1.0) and documentation are publicly available for download
from {\tt http://sigma-2.cesr.fr/spi/MGGPOD/}. In this paper we
provide an overview of the capabilities, functioning, and structure of
the MGGPOD package, and give examples of applications to past and
present gamma-ray missions.

\section{The MGGPOD Monte Carlo Simulation Suite}
\label{mggpod}

The MGGPOD Monte Carlo suite allows {\sl ab initio} simulations of
instrumental backgrounds -- including the many gamma-ray lines --
arising from interactions of the various radiation fields within the
instrument and spacecraft materials. It is possible to simulate both
prompt instrumental backgrounds, such as energy losses of cosmic-ray
particles and their secondaries, as well as delayed instrumental
backgrounds, which are due to the decay of radioactive isotopes
produced in nuclear interactions. Of course, MGGPOD can also be used
to study the response of gamma-ray instruments to astrophysical and
calibration sources. The MGGPOD suite is therefore an ideal Monte
Carlo tool for gamma-ray astronomy.
A detailed description of the physics of the MGGPOD Monte Carlo suite
can be found in \citet{weidenspointner_mggpod}. The documentation
available from the MGGPOD web site provides comprehensive practical
advice for users. First applications of MGGPOD have been presented in
\citet{weidenspointner_mggpod_seeon, weidenspointner_mggpod}.

\subsection{Capabilities and Functionalities}
\label{cap_func}

\begin{figure*}
\centering
\epsfig{file=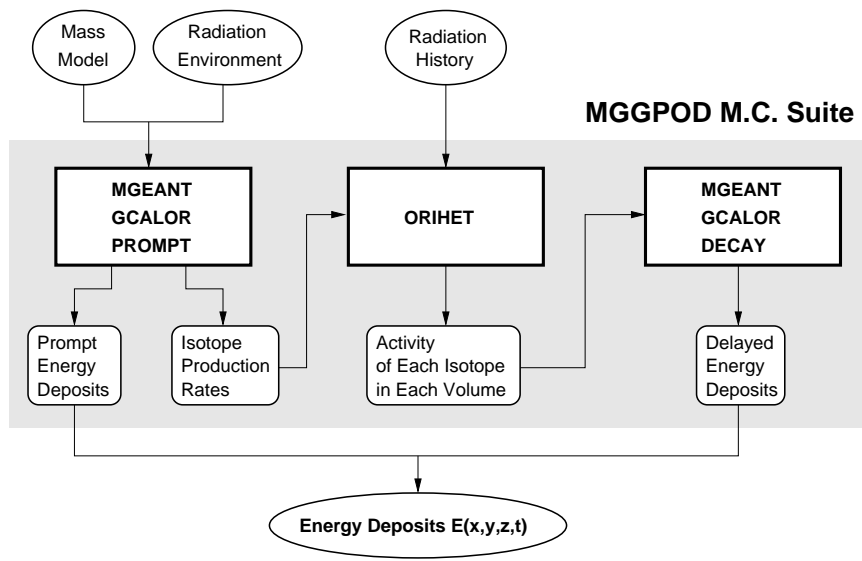,width=11.5cm}
\caption{A flow chart illustrating the overall
structure of the MGGPOD Monte Carlo simulation suite. The various
simulation packages (shown in boxes) and input and output files (shown
in ellipses and round-edged boxes) are explained in the
text\label{mggpod_flow_chart}}
\end{figure*}

MGGPOD is a suite of five closely integrated Monte Carlo packages,
namely {\bf MG}EANT, {\bf G}CALOR, {\bf P}ROMPT, {\bf O}RIHET, and
{\bf D}ECAY. The MGGPOD suite resulted from a combination of the
NASA/GSFC MGEANT \citep{Sturner00} and the University of Southampton
GGOD
\citep{Dean03} Monte Carlo codes, which 
we supplemented with the newly developed PROMPT package. All these
packages are based on the widely used GEANT Detector Description and
Simulation Tool (Version 3.21) created and supported at
CERN\footnote{see {\sl http://wwwinfo.cern.ch/asd/geant/}},
which is designed to simulate the passage of
elementary particles through an experimental setup.

In a nutshell, the capabilities and functions of the five
packages that constitute the MGGPOD suite are as follows:
\begin{itemize}
\item MGEANT is a multi-purpose simulation package that was created to
increase the versatility of the GEANT simulation tool. A modular,
``object oriented'' approach was pursued, allowing for rapid
prototyping of detector systems and easy generation of most of the
radiation fields relevant to gamma-ray astronomy. Within the MGGPOD
suite, MGEANT (i.e.\ GEANT) stores and transports all particles, and
treats electromagnetic interactions from about 10~keV to a few
TeV. MGEANT provides the option of using the GLECS
package\footnote{see {\sl
http://nis-www.lanl.gov/\symbol{126}mkippen/actsim/glecs/} by R.M.~Kippen}
to take into account the energy of bound electrons in Compton
scatterings. The MGEANT simulation package and a user manual are
available at a NASA/GSFC web site\footnote{see {\sl
http://lheawww.gsfc.nasa.gov/docs/gamcosray/legr/
mgeant/mgeant.html}}.
\item GCALOR 
\citep{Zeitnitz_Gabriel94} simulates
hadronic interactions down to 1~MeV for nucleons and charged pions and
down to thermal energies ($10^{-5}$~eV) for neutrons. Equally
important, this package\footnote{see {\sl
http://wswww.physik.uni-mainz.de/zeitnitz/gcalor/ gcalor.html}}
provides access to the energy deposits from all interactions as well
as to isotope production anywhere in the simulated setup.
\item PROMPT simulates prompt photon emission associated with the
de-excitation of excited nuclei produced by neutron capture, inelastic
neutron scattering, and spallation.
\item ORIHET, originally developed for the GGOD suite and improved for 
MGGPOD, calculates the build-up and decay of activity in any system
for which the nuclide production rates are known. Hence ORIHET can be
used to convert nuclide production rates, determined from simulations
of cosmic-ray irradiation, to decay rates. These are required input for
simulating the radioactive decays giving rise to delayed
background.
\item DECAY, again originally developed for GGOD and improved for our
purposes, enables MGGPOD to simulate radioactive decays.
\end{itemize}

\subsection{Structure}
\label{structure}

The overall structure of the MGGPOD package is illustrated in
Fig.~\ref{mggpod_flow_chart}. Depending on the simulated radiation
field or gamma-ray source distribution one or three steps, requiring
two or three input files, are needed to obtain the resulting energy
deposits in the detector system under study. In general, it is
advisable to simulate each component of the radiation environment
separately.
MGGPOD distinguishes two classes of radiation fields. Class~I
comprises radiation fields for which only prompt energy deposits are
of interest, such as celestial or laboratory gamma-ray sources or
cosmic-ray electrons. Class~II comprises radiation fields for which in
addition delayed energy deposits resulting from the activation of
radioactive isotopes need to be considered. Examples for Class~II
fields are cosmic-ray protons, or geomagnetically trapped protons.

For both of these classes, the simulation of the prompt energy
deposits requires two input files: a mass model, and a model of the
simulated radiation field. The mass model is a detailed computer
description of the experimental setup under study. It specifies the
geometrical structure of instrument and spacecraft, the atomic and/or
isotopic composition of materials, and sets parameters that influence
the transport of particles in different materials. Each component of
the radiation environment (and analogously for gamma-ray sources) to
which the instrument is exposed is characterized by three quantities:
the type of the incident particles, and their spectral and angular
distributions. The prompt energy deposits are written to an output
event file; in case of a Class~II radiation field there is an
additionial output file in which all the nuclei produced in hadronic
interactions are recorded.

To simulate delayed energy deposits (Class~II radiation field) two
additonal steps need to be taken. These require as input the time
history of the radiation field which is
responsible for the activation, and the previously calculated isotope
production rates.
Based on this information first the activity of each isotope produced in
each structural element of the mass model 
is determined. Then these activities are used to simulate the delayed
energy deposits due to radioactive decays in the instrument.

Combining prompt and delayed energy
deposits from each component of the radiation environment and
gamma-ray sources, it is possible to obtain the total energy deposited
in the system as a function of position and time.



\section{Practical Considerations}
\label{practic}

This section addresses some of the practical considerations a
potential user of MGGPOD might have. It is a synopsis of the
documentation available from the MGGPOD web site. 

Installation of the complete MGGPOD package (software and data files)
requires little more than 100~MB of disk space. The documentation
provides detailed installation instructions, including the
installation of the required CERNLIB and other libraries which are
necessary to build, but are not included in, the MGGPOD package.
The MGGPOD software is written in the FORTRAN~77 and C programming
languages; in addition there are a few C shell scripts. The source
code is completely open and transparent, allowing the user to adapt,
change, and improve the code. In fact, a few customizations and
adjustments of the code are inevitable when simulating different
instruments. These unavoidable changes are described in the
documentation. The format of both the event and activation output
files is FITS.
MGGPOD inherited the interactive display capabilities of (M)GEANT,
which are based on CERN's PAW++ package. This is a very convenient
feature when creating a mass model, or when specifying the parameters
of a radiation field (``beam'').

To facilitate the use of MGGPOD by novice users the release contains
examples for all simulation steps outlined in
Sec.~\ref{structure}. The examples include all input and output
files, and instructions on how to build and run the necessary
executables.

As described in Sec.~\ref{cap_func}, different modules of MGGPOD simulate
different physical processes which can be treated over different
energy ranges. The MGGPOD input file allows the user to define
low-energy cutoff values for the tracking of five different particle
types: photons, electrons, neutral and charged hadrons, and muons. We
recommend to use as default low-energy cutoff values 10~keV for
photons, electrons, and muons, $10^{-2}$~eV for neutral hadrons, and
1~MeV for charged hadrons. Experienced users can lower these
thresholds, in particular for photons.

For many applications of MGGPOD the computation time needed to
complete a simulation is an important consideration. The speed of
MGGPOD Monte Carlo simulations varies greatly with the type and energy
of the incident particle (which affect e.g.\ the number of secondary
particles that are produced and tracked), with the physical volume and mass
described in the mass model (but less with geometrical complexity),
and of course with the speed of the computer employed. 
As an example, for the TGRS and SPI mass models, using a 1.6~GHz CPU,
typically about $3 \times 10^{-3}$~CPU~seconds and $2 \times
10^{-2}$~CPU~seconds are needed to simulate the interactions of a
cosmic-ray proton, respectively. The difference in processing time is
mainly due to the fact that INTEGRAL/SPI is a much heavier and complex
instrument than WIND/TGRS. For both models the time needed to simulate
an incident gamma-ray photon, or a radioactive decay, is about ten
times shorter than the respective proton processing time. The
computation time for a simulation also depends on the desired
statistics in the result; better statistics require the simulation of
more particles and consequently more processing time. For SPI, about
$10^7$ photons, requiring a computation time of about 8~CPU~hours, in a
homogeneous beam covering the 19 Ge detectors are sufficient for
simulating the instrument response at a given energy. When simulating
the activation of INTEGRAL and SPI by cosmic-ray protons, $10^7$
protons, requiring a computation time of about 3~CPU~days (and
representing about 21~s of actual in-orbit irradiation), provide
acceptable statistics for calculating nuclide production rates, and
enough statistics in the detector count rate to obtain a crude
spectrum. Detailed studies of prompt background line production by
cosmic-ray protons require multiple simulation runs.

The MGGPOD web site also provides off-the-shelf model spectra for two
common radiation fields: cosmic-ray protons \citep[at solar minumum
and maximum, based on][]{moskalenko_galprop}, and the diffuse cosmic
X-ray and gamma-ray emission \citep[based on][]{Gruber_xrb_egb}. For
missions in low-Earth orbit albedo radiations (most importantly
neutrons and gamma rays) and geomagnetically trapped paricles need to
be considered in addition. The web site provides links to the ESA
Space Information System (SPENVIS) and the NASA Space Ionizing
Radiation Environments and Shielding Tools (SIREST), which can be used
to obtain radiation field models for missions in a low-Earth
environment. Users should be aware that the space environment is not
stable. Detailed simulations require that input particle spectra
reflect conditions in a given orbit and at a given phase of the solar
cycle \citep[including the polarity of the heliosphere for cosmic-ray
protons,][]{moskalenko_galprop}. Currently, GCALOR does not treat
hadronic interactions for alpha particles or heavier cosmic rays. If
diffuse cosmic X-rays are an important background component, the
uncertainty in the simulation result is usually dominated by
uncertainties in the photoelectric absorption in materials surrounding
the detectors as defined in the mass model.

\section{Applications of MGGPOD}
\label{applications}

\begin{figure}
\centering
\epsfig{file=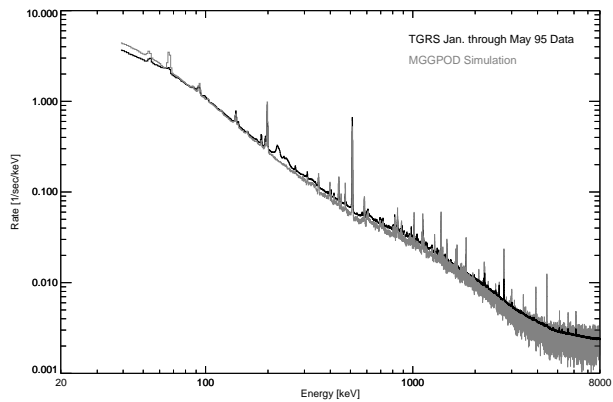,width=5.5cm,angle=90.}
\caption{A comparison of the January--May 1995 TGRS spectrum with a
MGGPOD simulation, taken from
\citet{weidenspointner_mggpod_seeon}. The broad features in the data 
between 210 and 260~keV are electronic artefacts \label{comp_tgrs}}
\end{figure}

The MGGPOD Monte Carlo suite has been, and is being, applied to model
the instrumental backgrounds of several gamma-ray missions. Using the
MGGPOD codes, very good agreement between the Monte Carlo results and
the actual data of the Transient Gamma-Ray Spectrometer (TGRS) on
board {\it Wind} \citep{Owens95} has been obtained, as discussed in
\citet{weidenspointner_mggpod_seeon, weidenspointner_mggpod}. As an
example, in Fig.~\ref{comp_tgrs} we show a comparison of MGGPOD
simulations with TGRS data.
%
First modelling results for the SPI Spectrometer on board the ESA
INTEGRAL observatory \citep{vedrenne_spi_instr} again yielded good
agreement, but also indicate remaining deficiencies with respect to
the production and thermalization of secondary neutrons in a massive
spacecraft and/or instrument
\citep{weidenspointner_mggpod_seeon}. Both of these instruments
operate in highly elliptical orbits above the Earth's radiation belts.

Recently, the MGGPOD suite has been applied to modelling the
instrumental background of the Reuven Ramaty High-Energy Solar
Spectroscopic Imager \citep[RHESSI, described in][]{Smith02}. This is
the first time MGGPOD is used for modelling the instrumental
background of an instrument in low-Earth orbit. 
This case is particularly difficult because activation during passages
through the South-Atlantic Anomaly (SAA) gives rise to a strong and
time-variable background component. Unlike all other components of
the radiation environment, the time-variability of the SAA-induced
backgrounds needs to be accounted for in the simulations. Preliminary
modelling results are presented by
\citet{wunderer_rhessi} at this workshop.


\section{Summary}

The MGGPOD Monte Carlo suite is an ideal tool for supporting the
various stages of gamma-ray astronomy missions, ranging from the
design, development, and performance prediction through calibration
and response generation to data reduction. The MGGPOD software and
documentation are publicly available for download at CESR. The package
has been, and is being, successfully applied to several past and
present gamma-ray missions.


\begin{thebibliography}{}
%
%
%
%
\bibitem[Dean et al.(2003)]{Dean03} Dean, A. J., et al. 2003, 
Space Sci.\ Rev.\ 105, 285
%
%
%
%
\bibitem[Gruber et al.(1999)]{Gruber_xrb_egb} Gruber, D.E., et al. 1999,
ApJ 520, 124
%
%
%
\bibitem[Moskalenko et al.(2002)]{moskalenko_galprop} Moskalenko, I.,
et al., 2002, ApJ 565, 280
%
\bibitem[Owens et al.(1995)]{Owens95} Owens, A., et al. 1995, 
Space Sci.\ Rev.\ 71, 273
%
\bibitem[Smith et al.(2002)]{Smith02} Smith, D.M., et al., Solar
Physics 210, 33, 2002
%
\bibitem[Sturner et al.(2000)]{Sturner00} Sturner, S., et al. 2000, {\sl The
Fifth Compton Symposium} (AIP 510), 814
%
\bibitem[Vedrenne et al.(2003)]{vedrenne_spi_instr} Vedrenne, G., et
al. 2003, A\&A 411, L63, 2003
%
\bibitem[Weidenspointner et al.(2003)]{weidenspointner_mggpod_seeon}
Weidenspointner, G., et al. 2003, New Astr.\ Rev.\ 48, 227
%
\bibitem[Weidenspointner et al.(2004)]{weidenspointner_mggpod}
Weidenspointner, G., et al. 2004, in preparation
%
\bibitem[Wunderer et al.(2004)]{wunderer_rhessi} Wunderer, C., et al.,
2004, these proceedings
%
\bibitem[Zeitnitz and Gabriel(1994)]{Zeitnitz_Gabriel94} Zeitnitz,
C., \& Gabriel, T. A. 1994, NIM~A 349, 106
%






\end{thebibliography}

\end{document}